\documentclass[nature,twocolumn]{revtex4-1}
\usepackage{graphicx}
\usepackage{dsfont, amsmath, amssymb}
\usepackage{bm}
\usepackage{natbib}
\usepackage{color}
\graphicspath{{images/}}

\newcommand{\bsw}{\boldsymbol{w}}

\newcommand{\magenta}[1]{{\color{black} #1}}
\newcommand{\cyan}[1]{{\color{black} #1}}

\begin{document}

\title{Long-range depth imaging using a single-photon detector array and non-local data fusion}

\author{Susan Chan$^1$, Abderrahim Halimi$^1$, Feng Zhu$^1$, Istvan Gyongy$^2$, Robert K. Henderson$^2$, Richard Bowman$^3$, \magenta{Steve} McLaughlin$^1$, Gerald S. Buller$^1$ and Jonathan Leach$^1$}

\affiliation{$^1$ School of Engineering and Physical Sciences, Heriot-Watt University, Edinburgh, EH14 4AS, UK}
\affiliation{$^2$ Institute for Integrated Micro and Nano Systems, The University of Edinburgh, Edinburgh, EH9 3JL, UK}
\affiliation{$^3$ Department of Physics, University of Bath, Bath, BA2 7AY, UK}

\begin{abstract}
The ability to measure and record high-resolution depth images at \cyan{long stand-off distances} is important for a wide range of applications, including connected and automotive vehicles, defense and security, and agriculture and mining.  \cyan{In} LIDAR (light detection and ranging) \cyan{applications}, single-photon sensitive \cyan{detection is an emerging approach, offering high} sensitivity to light and picosecond temporal resolution, \cyan{and consequently excellent surface-to-surface resolution.  The use of large format CMOS single-photon detector arrays provides high spatial resolution and allows the timing information to be acquired simultaneously across many pixels.}  In this work, we combine state-of-the-art single-photon detector array technology with non-local data fusion to generate high resolution three-dimensional depth information of long-range targets.  The system is based on a visible pulsed illumination system at 670~nm and a 240~$\times$ 320 pixel array sensor, achieving sub-centimeter precision in all three spatial dimensions at a distance of 150 meters.  The non-local data fusion combines information from an optical image with sparse sampling of the single-photon array data, providing accurate depth information \cyan{at low signature regions of the target}.
\end{abstract}

\maketitle

\cyan{Imaging} technology \cyan{capable of measuring} high resolution three-dimensional depth information has developed significantly in recent years.  \cyan{Such 3D imaging technology is required in a range of emerging application areas:  for example}, the gaming industry requires accurate high-speed player position information \cite{Kolb:2010fu}; \cyan{the defense sector requires} long-range target identification \cyan{for several scenarios} \cite{Shen:2018im, Tobin:2018cq}; and in the automotive sector, situational awareness technology will play a key role in the future of connected and autonomous vehicles \cite{Schwarz:2010jh, Niclass:2015jo, Itzler:2017vh}.  In parallel with the technological advances in 3D imaging hardware, computational image processing has been shown to be extremely powerful \cite{Altmann:2018hn}.  Artificial neural networks can be trained to identify images in low-light levels \cite{Nasrabadi:2007ea, Niu:2018fl}, and optimization procedures, based on prior information, can be used to de-noise, upscale, and enhance 3D images \cite{Altantawy:2017hn, Ren:2018jj}.  Whilst each application has differing performance \cyan{requirements}, ultimately the future of three-dimensional imaging will rely on \cyan{a combination of} state-of-the-art \cyan{camera} technology \cyan{and} advanced image processing methods \cite{Kirmani:2014fn, Howland:2011cp}.

LIDAR (light detection and ranging) is a \cyan{commonly used} method for determining the distance of an object based on the \cyan{time of flight of the optical signal returned from the target} \cite{Jaboyedoff:2010bj, Simard:2011de, Dong:2303937}.  Here a pulsed illumination source is used in combination with a \cyan{single-photon} detector.  Once the return signal is measured, the \cyan{round-trip} time \cyan{of the return signal is estimated and the distance to the target can be deduced}.  \cyan{Many} LIDAR \cyan{systems operate} in a point-like fashion, where the distance to a point on an object is measured.  If full three-dimensional \cyan{imaging} is required, the source is scanned over the object and the depth information is built up \cyan{pixel by pixel}.

The \cyan{maximum range of} a LIDAR system \cyan{is limited by a number of factors, including the optical power levels and the geometry of the receive channel (aperture diameter\magenta{, etc.}).  A major factor limiting maximum range is the type of optical detection scheme, with single-photon detection being considered a potential option, due to its high sensitivity and excellent surface-to-surface resolution.  For example, single-photon detection has been used in demonstrations at over 10~km range in daylight conditions \magenta{\cite{McCarthy:2009by, McCarthy:2013bl, McCarthy:2013cq, Pawlikowska:2017hk, Kang:2018dm}}, allowing detailed reconstruction of non-cooperative targets.  Superconducting nanowire single-photon detectors have shown good performance in LIDAR and depth profiling applications \cite{Warburton:2007gs, McCarthy:2013cq}, however their low operating temperatures (i.e. $<$~4~K) has limited their use in practical applications.  SPAD detectors have shown great potential for LIDAR applications being operated at, or near, room temperature, as well as being highly sensitive and exhibiting low jitter (typically $<$~100~ps).  As examples, SPADs have been used in a range of LIDAR and depth imaging demonstrations:  1~km imaging using Si-SPADs \cite{McCarthy:2009by}, imaging at 1550~nm wavelength with InGaAs/InP SPADs \cite{McCarthy:2013bl}, and 10~km imaging at $\lambda$~=~1150~nm \cite{Pawlikowska:2017hk}.  In each case, detailed three-dimensional reconstruction was possible.}  Single-photon counting LIDAR \cyan{has been demonstrated in several applications}, including remote sensing for geodesy \cite{Glennie:2013eg}\cyan{, for target identification in clutter \cite{Henriksson:2017ju, Tobin:2018cq},} and \cyan{for airborne analysis of} forestry \cite{Swatantran:2016dt} \cyan{and multispectral analysis of arboreal physiological parameters \cite{Wallace:2014fp}}.

Over the past several years, single-photon sensitive detectors have been used to observe laser propagation in air \cite{Velten:2013cd, Gariepy:2015ca}, to detect objects hidden from the line-of-sight \cite{Velten:2012ik, Gariepy:2015gi, Chan:2017ku, Laurenzis:2017el, Caramazza:2018je}, and to image in the presence of scattering media \cite{Laurenzis:2012hn, Maccarone:2015de, Halimi:2017ha, Satat:2017jh, Satat:2018co, Zhu:2017hb, Tobin:2018ed}.  In particular, progress is being made in the development and implementation of SPAD detectors in the form of dense pixel arrays \cite{Albota:ve, Niclass:2005cz, Zappa:2007fi, Shin:2016fy}.  Such detectors take advantage of metal-oxide-semiconductor (CMOS) technology, so that an array of SPADs can be fabricated and integrated onto a small chip \cite{Dutton:2014bh, Gyongy:2018ip}.  SPAD arrays consist of large number of pixels that provide high spatial resolution while \cyan{maintaining single-photon sensitivity and picosecond timing resolution} \cite{Gyongy:2018hz}.  They are ideal candidates for three-dimensional ranging and imaging due to their high-spatial resolution and temporal characteristics \magenta{\cite{Ren:2018jj}}.

In this work we demonstrate a high-resolution three-dimensional imaging system based on a silicon CMOS SPAD array and an active illumination system.  The precision of the system is measured to be less than a centimeter in all three spatial dimensions at a stand-off distance of 150~meters.  We then combine our results with a non-local data fusion algorithm \cite{Buades:2005ib, Dabov:ub, Salmon:2014ge} that can \cyan{complete} missing depth information by exploiting information from a co-registered optical RGB image taken with a digital camera.  We use weighted color and depth information to fill in the \cyan{missing information} \cite{Park:2014cc, Yang:2014ho}, and assume that close regions have local correlations in depth \cite{Rudin:1992kn, Iordache:2012ck, Lanaras:2017wx}.  This ability to fill in missing information is extremely powerful as it can enable large areas of depth information to be gathered from a small subset of data.  Our results demonstrate the capability of such sensors at measuring depth at long distances and illustrate the potential for extremely high-resolution imaging at distance.

{\subsection{Results}}\label{subsec:results}

\begin{figure*}[htbp]
\centering
\includegraphics[width=2\columnwidth]{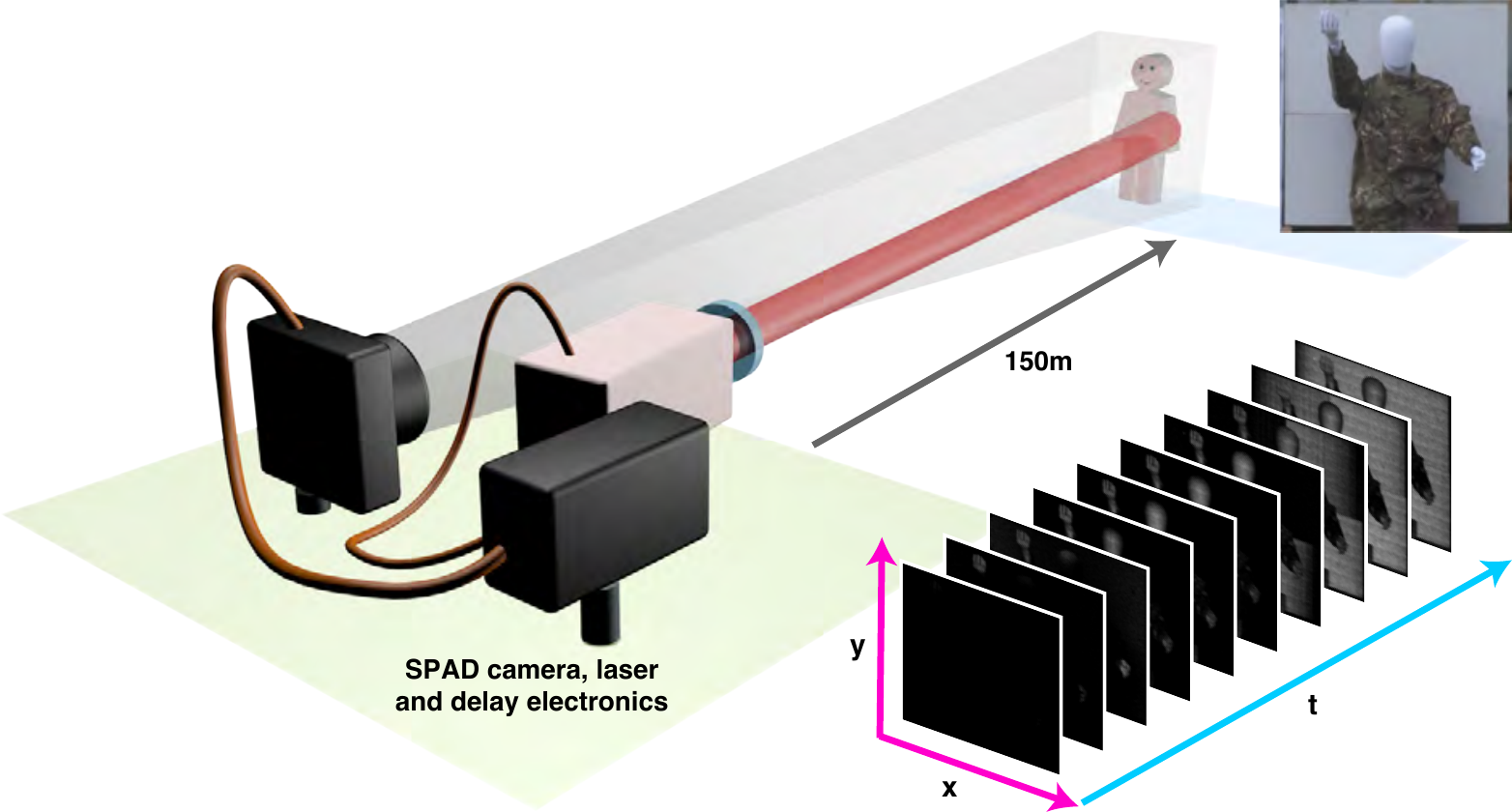}
\caption{Experimental setup for SPAD-based time-gated image sensing.  The distance from the SPAD camera and laser to the target \magenta{is} 150 m.  The inset photograph is \magenta{the co-registered RGB image captured and used for the non-local fusion image processing}.  The series of images represent \magenta{sample preprocessed images corresponding to ten consecutive gate positions}.  Depth information is gained by scanning the location of the gate.}
\label{fig:concept}
\end{figure*}
{\textbf{Experimental setup.}}
The system is configured in a bistatic setup comprising three parts:  the transmitter, the receiver and the associated system control components as shown in Fig. \ref{fig:concept} (see Methods for further details).  The Single-Photon Counting Imager (SPCImager) is a SPAD-based time-gated image sensor implemented in 0.13~$\mu$m silicon CMOS with 8~$\mu$m pixel pitch and 26.8$\%$ fill factor.  It operates over a wavelength range of around 350~nm~-~1000~nm.  This is one of the largest-format SPAD \cyan{detector arrays produced} to date.  It is a high frame-rate imager (up to 10~kfps) that offers high spatial resolution with its array of 240 by 320 pixels, and picosecond timing resolution with its time-gating capability.  The rise time of the electronic gate is of the order of tens of picoseconds.  The minimum width of this gate is 18~ns.  

The stand-off distance from the active imaging system to the \magenta{target} scene is $\sim$150~m.  Data for a three-dimensional image is collected by scanning the laser beam over the target surfaces in a 20 by 20 grid and recording the respective signal returning from the scene for a single gate setting (see Supplementary Information for an illustration of the scanning procedure), and then repeating this process for a series of different gate settings.  The laser illumination spot is defocused such that it approximately covers a 50 by 50 pixel area.  This size of illumination is chosen as a good compromise between the number of scan positions and the signal return from the target.  The exposure time per scan position is 215~$\mu$s, and 256 bit planes are added together per scan position (chosen empirically).  We measure an intensity related to the reflectivity of the objects in the target scene.\\
\begin{figure*}[htbp]
\centering
\includegraphics[width=2\columnwidth]{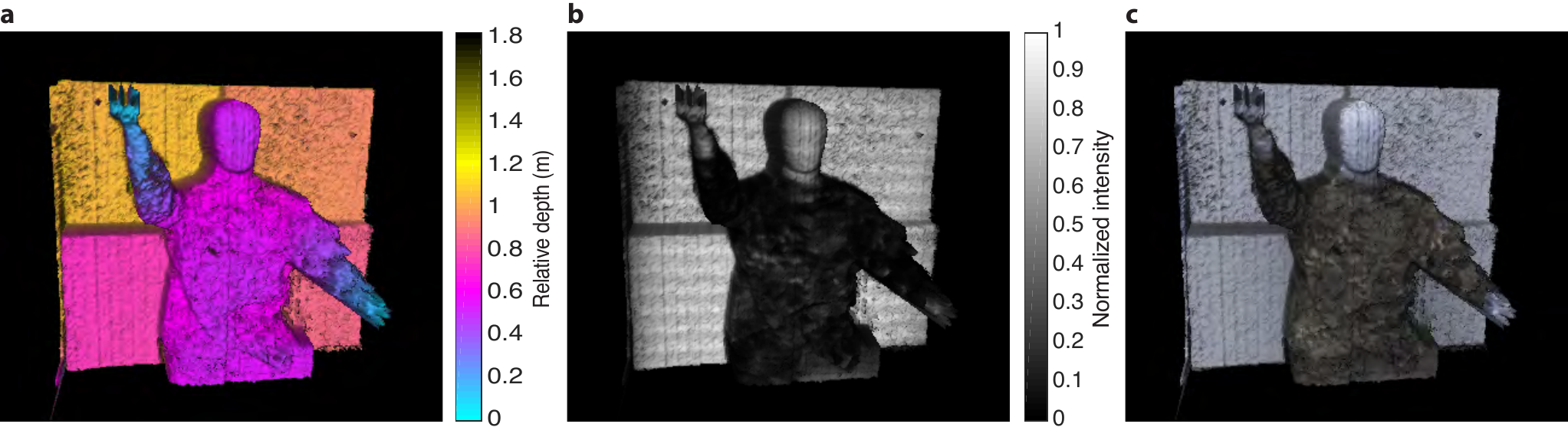}
\caption{Three-dimensional reconstruction of the target scene at 150~m distance.  (a) Retrieved depth information.  (b) The intensity information from the SPAD overlaid on top of the retrieved depth information.  (c) The intensity information from a DSLR camera overlaid on top of the retrieved depth information.}
\label{fig:fullresult}
\end{figure*}
{\textbf{Depth imaging at 150~m.}}
After the data acquisition and initial processing, we have a 240 by 320 by 51 three-dimensional data cube.  For each pixel in the 240 by 320 array, we can expect the recorded intensity to be approximately zero when the target depth is outside the time gate of the SPAD camera.  Otherwise, it is a non-zero constant.  To establish the target depth and intensity information at each pixel, we apply the following fitting function to the observed data $y_{n,k}$, the number of photon counts measured by pixel $n$ for a depth sample $k$ (see Supplementary Information for sample fit):
\begin{equation}
s_{n,k} = \frac{r_n}{2} \left\lbrace 1 + \textrm{erf} \left[\frac{k- d_n}{h} \right]\right\rbrace + b_{n}
\label{eqt:erf}
\end{equation}
where erf$(.)$ denotes the error function, $h\geq 0$ is a fixed impulse response (IR) intrinsic parameter that represents the width of the leading edge, and $d_{n}\geq 0$ and $r_{n}\geq 0$ are related to the target's depth and intensity respectively.  Here, we assume the absence of background $b_{n}=0$, $\forall$~$n$, since the noise has been removed at the preprocessing stage.  Although we utilize the erf$(.)$ function, other kinds of bounded monotonic increasing functions in the real number domain, such as arctan$(.)$, may also be considered.

Figure~\ref{fig:fullresult} shows 3D reconstructions of the target scene obtained by combining the depth and intensity information retrieved using the non-linear least squares fitting method.  The results have been cropped to the scene (228 by 228 pixels).  We apply a correction to the retrieved depth profile to account for the fall time mismatch of the electronic gate of the SPAD sensor.  In order to determine the range and surface-to-surface resolution of our system, we analyze a patch of 25 by 25 pixels for each panel of the depth board target.  We calculate a standard deviation and hence a range resolution of 0.96~cm, 0.82~cm, 0.89~cm, and 0.59~cm for the top-left, top-right, bottom-right and bottom-left panels respectively.  We also calculate a mean difference of 9.26~cm, 8.79~cm, 11.05~cm, and 29.10~cm between each pair of adjacent boards (clockwise from the top-left panel).  These numbers should be compared to 10 cm, 10 cm, 10 cm, and 30 cm.\\

\begin{figure}[htbp]
\centering
\includegraphics[width=1\columnwidth]{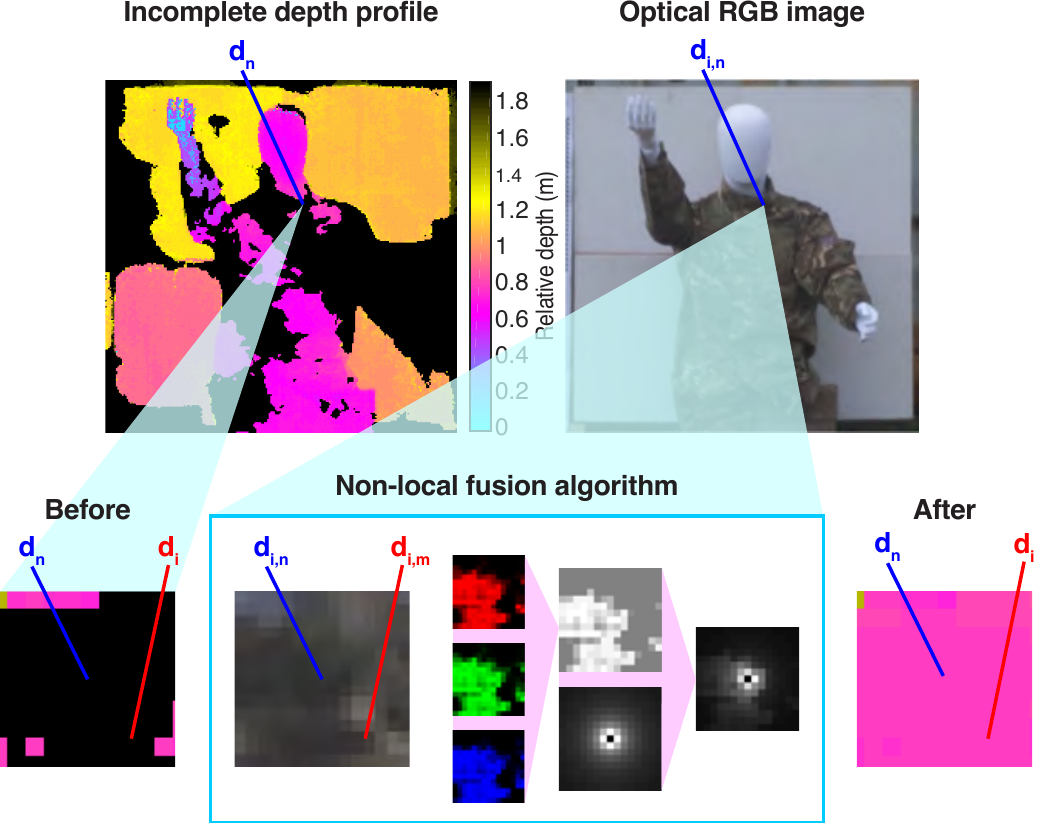}
\caption{Non-local data fusion.  For every pixel in the optical image, differences in RGB color values are calculated, summed and averaged for that pixel and every other pixel in a predefined field.  This is then weighted by the normalized vector distance between the pixels to obtain a weight for the regularization function to fill in the depth information missing from the scene.}
\label{fig:algorithm}
\end{figure}
\begin{figure}[htbp]
\centering
\includegraphics[width=1\columnwidth]{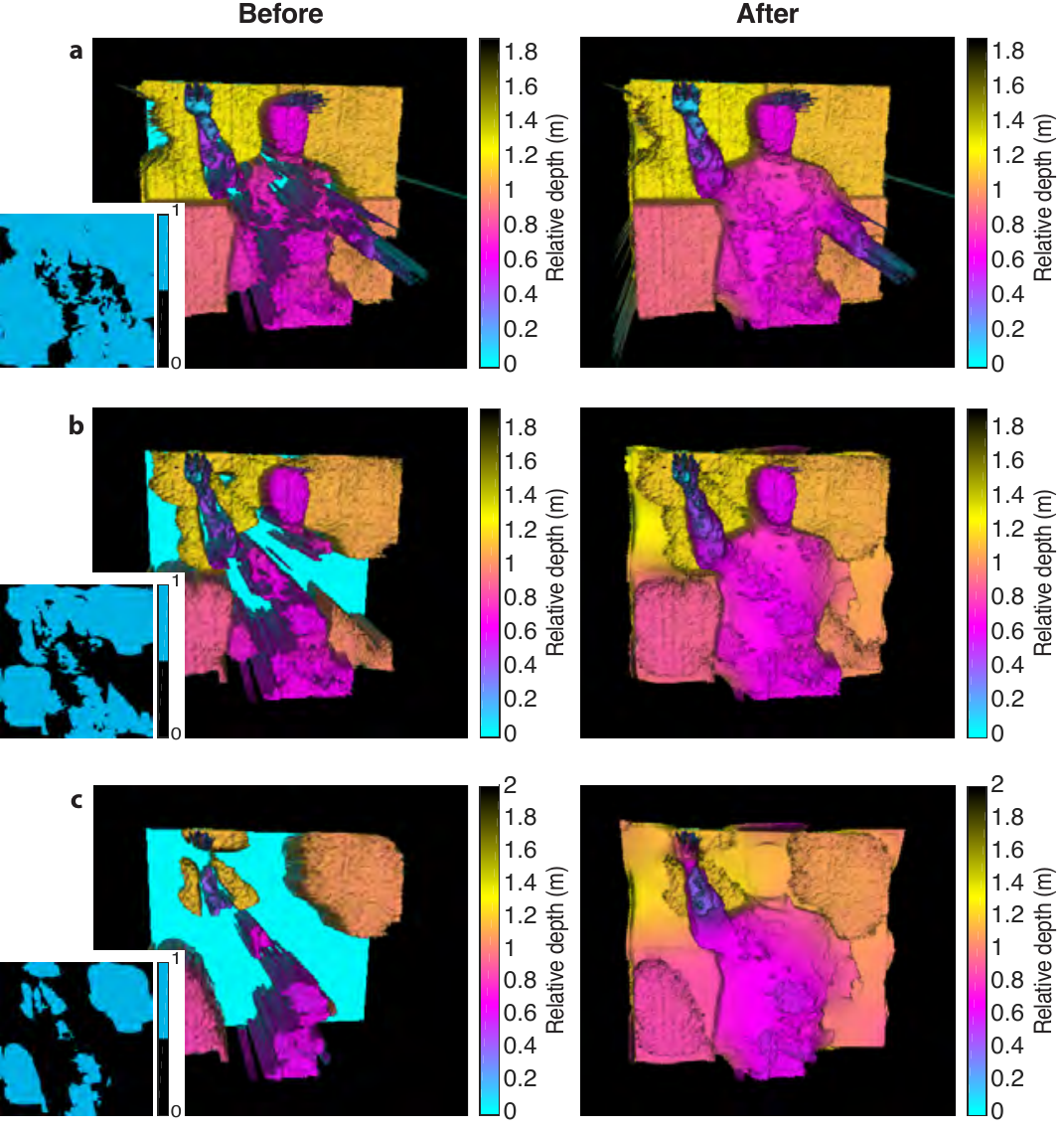}
\caption{3D reconstruction of the target scene using (a) 25$\%$, (b) 10$\%$, and (c) 5$\%$ of the scanned positions before and after non-local data fusion.}
\label{fig:class_nlf}
\end{figure}
%
%%%%%%%%%%
{\textbf{Non-local fusion-based image processing.}}
LIDAR data acquisition is performed by scanning several beam locations at each depth.  This can lead to a high acquisition time, creating a bottleneck for the deployment of such a system in real-life applications.  It is therefore important to deal with this challenge, and there are two ways to do this:  (i) reduce the number of bit planes summed at each depth position (i.e. reduce the dwell time), and (ii) perform compressed sensing by considering under-sampled and random beam scans. Both solutions lead to sparse pixels containing \cyan{fewer} photons and, consequently, having a higher noise level.  Under these conditions, using the noise statistics for parameter estimation is necessary but not sufficient to obtain good parameter estimates for Eq.\eqref{eqt:erf}.  In addition\cyan{,} rapidly scanning the laser across a scene can lead to areas with little or no depth information.  This limitation can be overcome by carefully considering prior information.  For example, it is known that images exhibit strong spatial correlations, i.e.~images tend to be smooth and composed mainly of low spatial frequencies, and the same can be said for depth.  This prior information can be used by considering regularization terms that account for correlations in the estimated depth and intensity images.

The core of our algorithm is to make use of a co-registered optical image, which has complete RGB intensity information for every pixel, and prior knowledge about the object to help fill in missing information in the depth profile.  This missing information arises as we do not consider all of the measured scan positions, but we simulate a rapid scan thus losing depth information about certain pixels.  As a result, we have pixels in the image where we have complete depth information and pixels where we have no depth information.  We do, however, have complete RGB color information provided by the optical image.  Our algorithm makes use of a core assumption:  the depths of two objects are strongly correlated \cyan{with} the similarity of their RGB values and a function of the distance between them, i.e. two objects that are close by and have similar colors will also be at similar depths.  This means that there will be a correlation in the depth of pixels with the same color, and this correlation will be stronger the closer those pixels are.   This assumption is extremely powerful as it enables us to fill in the depth information about an object when we only have knowledge about the RGB \cyan{value.  Note that similar assumptions are commonly adopted for the restoration of depth maps, as in \cite{Rapp:2017bc, Yang:2014ho}.}

In our non-local optimization, we consider four terms.  The first term ensures agreement between our data and a Poisson statistical noise model while the second imposes non-negativity on the estimated parameter values.  Of particular interest are the third and fourth terms which are our regularization terms for the target's depth and intensity respectively.  Each of these regularization terms is the sum of weighted differences between each pixel and other pixels located in a predefined field \cyan{of nearby pixels}.

To obtain the weights for the intensity regularization function, we first take the optical image and rescale this to the same resolution of the depth image (228 by 228 pixels).  For every pixel in the image, we then calculate the difference between \cyan{the RGB value of} that pixel and \cyan{the RGB value of each nearby pixel} in the predefined field, for each of the three RGB color channels (the field is defined to be a 15 by 15 pixel square centered on each starting pixel).  The differences of the three RGB channels are summed and averaged for the pixel leading to a weight value for each pixel and each direction.  \cyan{These weights are only calculated between each pixel and the 15 by 15 square region around it, in order to improve the efficiency of the calculation.}

This approach enforces small weighted differences between similar pixels in the image while preserving sharp edges \cyan{by considering an $\ell_1$ norm to compute pixel differences (see Eq.(11) in Supplementary Materials).  Note that other commonly used transformations can be applied to the RGB image before computing the weights, such as YUV or YCrCb \cite{Lebrun:2013ic},  and this can be easily included in the proposed algorithm.}  Each 15 by 15 matrix of differences is reshaped to a 225 by 1 vector, such that we have a 228 by 228 by 225 array containing all weighted differences for all pixels.  This matrix $\bsw$ \magenta{has} elements $w_{nm}$ where the index $n$ corresponds to the pixel of interest, and $m$ corresponds to the direction from that pixel (see Supplementary Information for details).  For the regularization of the depth, we take each intensity weight in the matrix and further weight this with the \magenta{normalized} vector distance between each pixel $n$ and the corresponding $m$th pixel.  The non-local fusion image processing is summarized in Figure~\ref{fig:algorithm}.

Figure \ref{fig:class_nlf} shows the results of the non-local fusion algorithm.  These data correspond to 25$\%$, 10$\%$\magenta{,} and 5$\%$ of scan positions, which \magenta{relate then to} 83.7$\%$, 61.8$\%$\magenta{,} and 35.8$\%$ of pixels respectively.   \magenta{Again, we have applied a correction to the retrieved depth profile to account for the fall time mismatch of the SPAD sensor gate.}  Significant improvement in the three-dimensional image is observed after data processing, even in the case where 5$\%$ of the scan positions are used.  Features such as the mannequin's arm and head can be recognized in the processed data, even when they are not present in the original depth data set.  This shows the clear advantage of the proposed algorithm in data reconstruction \cyan{with} an extremely reduced number of scan positions\cyan{, which could correspond to a significantly reduced acquisition time}.\\

%%%%%%%%%%

{\subsection{Discussion}}\label{subsec:discussion}

The system has performed exceptionally well at recording three-dimensional depth information at a distance of 150~m.  We are able to record depth data with a standard deviation less than one centimeter using a gate separation of 0.25~ns (7.5~cm).  Lower standard deviations will be possible if smaller gate \magenta{separations} are used.

The drawback of the system in its current form is the time required to generate first an image and then scan the gate position.  The time taken to collect all the data was around 2 hours. In principle, the same data could be collected in a fraction of the time.  The total acquisition time for each of the images in the 20 by 20 scan is only 215~$\mu$s.  This means that the total exposure time for all 400 images is only 100~ms.  We should therefore be able to get an intensity image at a frame rate of around 10 frames per second.  It would then take a \cyan{total of} 5 seconds to generate the 3D data cube, which contains all the 3D information.  \magenta{Additionally, the next generation of SPAD detector arrays with TCSPC capabilities will provide the ability to capture 3D depth data at higher acquisition rates \cite{Henderson:2018wm}.}

The system offers potential detection in low visibility and low light level environments.  In particular, the time-gated imaging approach may be extended to degraded visual environments in which a scattering medium, e.g. dust, fog, rain, smoke, and snow, is present.  By not activating the sensor until the outgoing photon has penetrated the medium to reach the object of interest, the main advantage that this approach provides is the capability to reject light that has been backscattered from everywhere else in the scene and, as a result, obtain an image that would otherwise be severely degraded by early backscattered photons.\\

\subsection{Methods}\label{subsec:methods}

\begin{figure}[htbp]
\centering
\includegraphics[width=1\columnwidth]{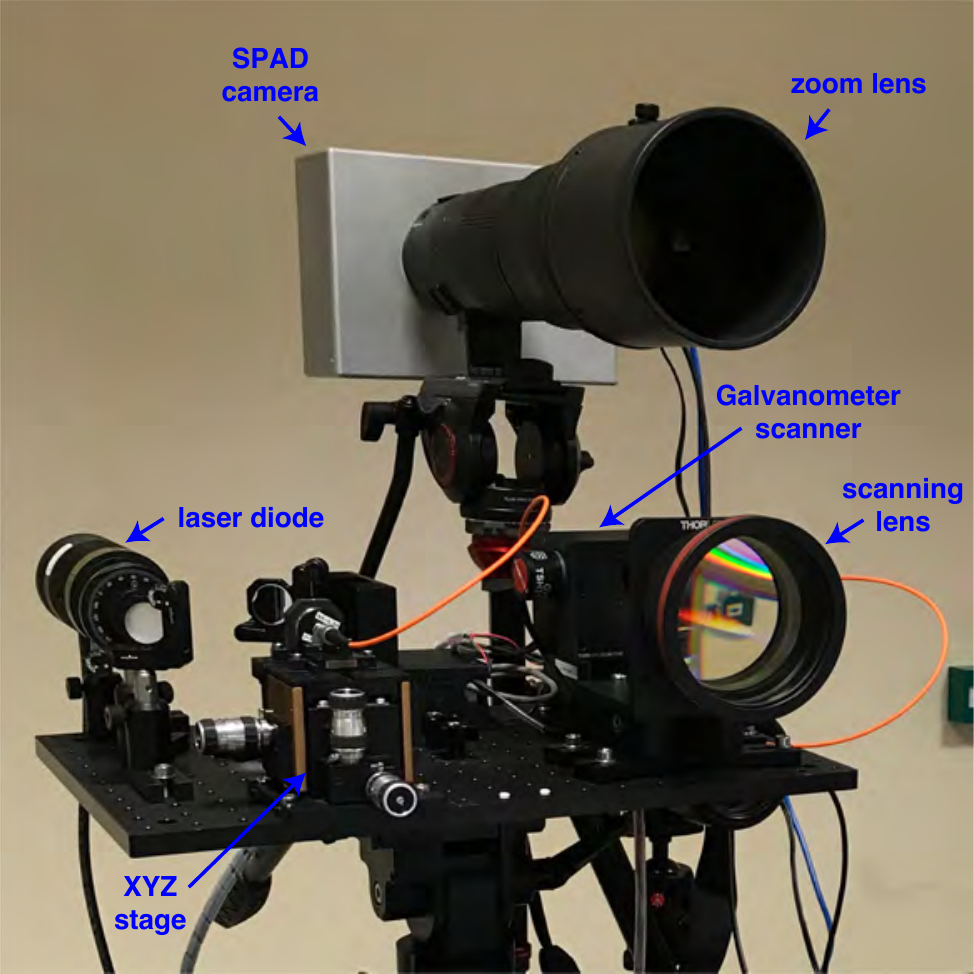}
\caption{Photograph of the scanning pulsed illumination system and SPAD camera.}
\label{fig:system}
\end{figure}
\begin{figure}[htbp]
\centering
\includegraphics[width=1\columnwidth]{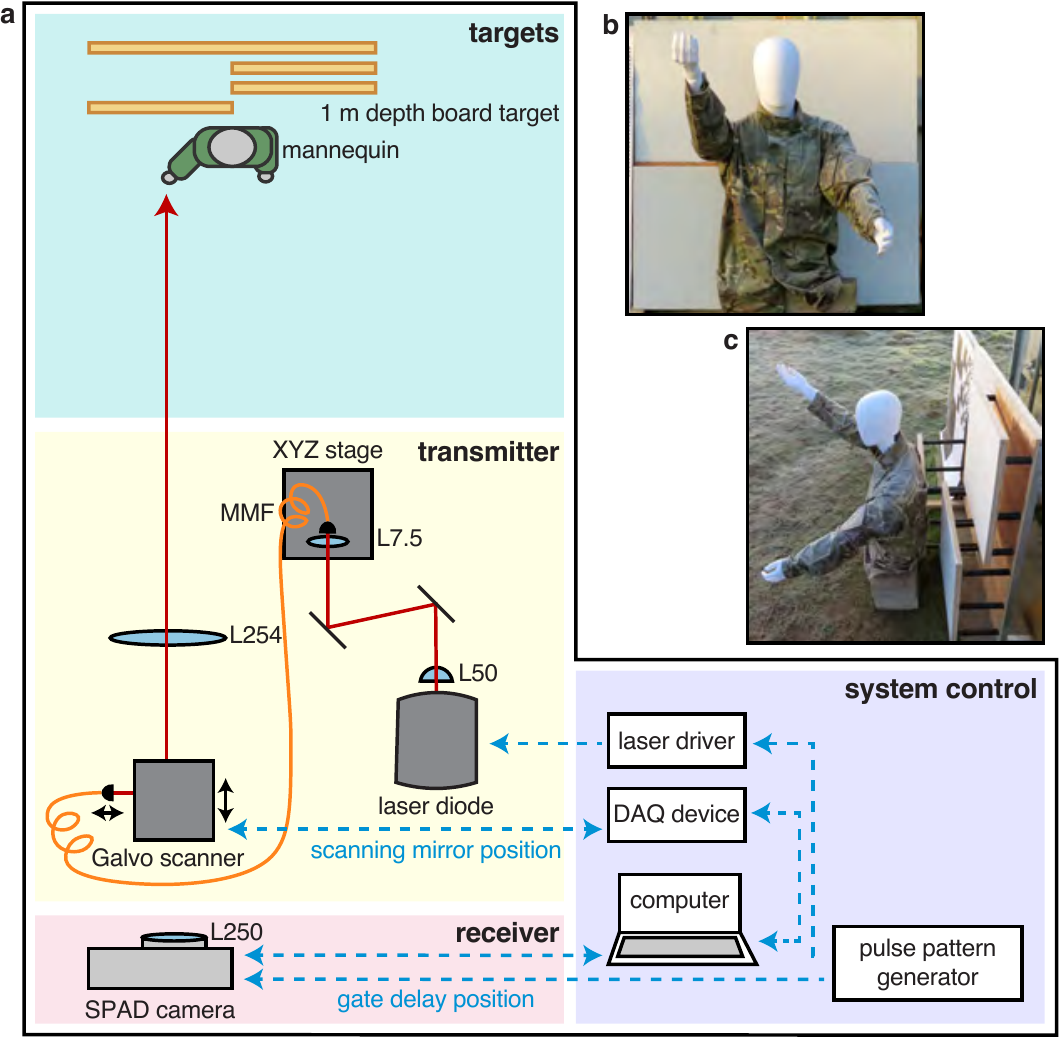}
\caption{Imaging system and target scene.  (a) shows a schematic of the transmitter, the receiver and the associated system control components.  (b) front-profile photograph of the mannequin and the 1~m depth board target located at 150~m stand-off distance.  Lighting conditions are as in the experiment. (c) side-profile photograph of the targets.}
\label{fig:setup}
\end{figure}
{\textbf{Details of experimental setup.}}
\magenta{A photograph of the system is shown in Fig. \ref{fig:system}.}  The transmitter (see Fig.~\ref{fig:setup}) is a scanning pulsed illumination system mounted on an M6-threaded 300~mm by 450~mm aluminum breadboard attached to a Kessler K-Pod heavy-duty tripod with Hercules 2.0 pan-and-tilt tripod head.  The light beam from a visible picosecond pulsed laser diode head (PicoQuant LDH-P-C-670M, 671~nm peak wavelength, 15~MHz repetition rate, 40~mW average power) is reshaped using a 1-inch N-BK7 plano-convex cylindrical lens (f~=~50~mm) before being coupled, using two 1-inch broadband dielectric mirrors, into a 50~$\mu$m core diameter multimode fiber patch cable (0.22~NA, 1~m).  The fiber input is mounted onto a manual XYZ flexure stage (Elliot Scientific) with a microscope objective lens (0.30~NA, f~=~7.5~mm); the output is mounted onto a 30~mm cage mount system.  The light exiting the fiber (average power $\sim$19~mW) is launched into the motor and mirror assembly of a dual-axis scanning Galvanometer mirror positioning system (Thorlabs GVS012/M), after which it passes through an f-theta scanning lens (EFL~=~254~mm).  The object distance to the scanning lens is changed by adjusting the position of the fiber cage mount and the position of the Galvanometer scanner, which is mounted on a dovetail optical rail.  This allows us to control the size of the illuminating spot at the target.  There is a trade-off when selecting the size of the illumination at the target.  On the one hand, it is ideal to have a large spot illuminate the target as images can be formed quickly without the need to scan.  On the other hand, larger return signals can be achieved if the spot is focused at the target.

The receiver (see Fig.~\ref{fig:setup}) consists of a tripod-mounted commercial zoom lens (Nikon AF-S NIKKOR 200~-~400 mm f/4G ED-IF VR) onto which a SPAD camera is mounted.  For the experiment, the focal length of the lens is set to f = 250~mm and the aperture is set to f/4. Inside the aluminum camera housing, between the lens and the SPAD sensor, we place a 1.5~nm bandpass filter to limit the background light and accept only the incoming light that matches the peak wavelength of the laser source.  An Opal Kelly XEM6310-LX45 FPGA integration module provides the electronic readout from the SPAD sensor chip via USB 3.0 to a control computer; see Fig.~\ref{fig:setup}.  The computer also connects to the Galvanometer scanning system through a DAQ device (National Instruments NI USB-6211), controlling the range and number of steps that the scanner takes.  A laser driver (PicoQuant PDL 800-D) controls the pulse repetition rate and output power of the laser diode.  A pulse pattern generator (Keysight 83114A) triggers the laser and SPAD camera, allowing synchronization between the return photons and the electronic gate.  It also controls scanning of the temporal delay of the gate.

The cyan section in Fig.~\ref{fig:setup} shows a birds-eye view of the composite elements of the target scene located at 150 meters from the system.  The photographs (inset in Fig.~\ref{fig:setup}) show the front and side \magenta{profiles} of the target objects.  In the scene, a mannequin is dressed in camouflage military clothing with its arms positioned to vary the range of depths that is measured across the scene.  Behind the mannequin, we place a 1~m square wooden depth target comprising four 0.5~m square panels at incremental depths of 0~cm, 10~cm, 20~cm\magenta{,} and 30~cm \magenta{(measured clockwise from the top-left panel)}.\\

{\textbf{SPAD-based time-gated image sensing.}}
The system performs time-gated imaging using a pulsed picosecond light source and the SPCImager operated in gated mode.  \cyan{When the laser diode emits a pulse, the SPAD camera is triggered to acquire data for the gate duration of 18~ns.}
By introducing a temporal delay to the \cyan{trigger signal from the laser diode}, we synchronize the \cyan{return} light \cyan{signal} with the electronic gate of the camera \cyan{in order to image photons returning from a pre-determined range}.  The SPCImager is a quanta image sensor that gives a binary output indicating whether or not a photon is present.  A bit plane is a single exposure of the camera's sensor and is a 240 by 320 array of 0s and 1s.  Multiple bit planes are required to build up an image with any grayscale.\\

{\textbf{Data acquisition.}}
Time-gating of the camera enables sectioning of the target scene.  The gate width is set to 18~ns, i.e. all photons arriving within 18~ns of the temporal delay are captured.  However, the rise time of the gate is very fast, so objects can come in and out of the captured image very clearly, depending on the location of the gate.  That is to say, the gate can be delayed so that the return light from certain objects are inside the imaging gate and are seen by the camera, but the return light from other objects are located outside of the gate and therefore rejected.   Three-dimensional information about a target is gained by \cyan{temporally} scanning the gate across the object and measuring the precise time at which the object enters the gate.  This time is then converted into distance.  In this experiment, images at 51 \cyan{temporal} gate positions are recorded.  The spacing between each image is chosen to be 0.25~ns, corresponding to a separation of approximately 7.5~cm in distance.

The depth and intensity information of the target surface is obtained by first generating a three-dimensional data cube consisting of intensity images taken at different depths.  The depth at which these images are captured is determined by the location of the gate.  Each image at a particular depth is constructed from individual image frames that correspond to around 400 different laser illumination positions.  Each of these frames record the information from the target scene and noise, which is a combination of the background light and dark counts from the sensor.   We find that the noise associated with background light does not change significantly from image to image.  This is in contrast to the noise associated with dark counts; pixels with high dark count rates can have a high variation from frame to frame.   There are, however, only a few pixels with very high dark count rates (more that 90$\%$ of the pixels have a dark count rate of less than 10~kHz), and this can be accounted for easily.  We deal with each of these sources of noise separately (see Supplementary Information).\\

{\subsection{References}}\label{subsec:ref}

\bibliographystyle{naturemag}
%\bibliography{Depth}

{\subsection{Acknowledgements}}\label{subsec:acknowledgements}
\magenta{This work was funded by the Engineering and Physical Sciences Research Council (EPSRC, UK) through the Quantum Technology Hub in Quantum Enhanced Imaging (EP/M01326X/1),
by the European Research Council TOTALPHOTON grant through the EU's Seventh Framework Program (FP/2007-2013)/ERC GA 339747,
by the UK Royal Academy of Engineering under the Research Fellowship Scheme (RF/201718/17128),
and by the Defense Science and Technology Laboratory's \cyan{Defense and Security Accelerator (DASA)}.
The assistance of Phil Soan in organizing and coordinating the Imaging Through Obscurants (ITO) Fog and Smoke Trial at the DSTL Porton Down laser range is gratefully acknowledged.  The authors appreciate the support of STMicroelectronics who fabricated the SPCImager.}\\

{\subsection{Author contributions}}\label{subsec:authors}
S.C. and J.L. performed the experiments and drafted the manuscript.  S.C., A.H., S.M.\magenta{,} and F.Z. performed data analysis.  R.H. designed the CMOS SPAD pixel architecture and I.G. provided support with the SPAD array.  R.B. wrote the LabView VI that allows 3D visualization of an object by overlaying its depth profile with an intensity profile.  G.B. and J.L. led the project.  All authors contributed to scientific discussions and to the final manuscript.\\

{\subsection{Additional information}}
The authors declare no competing interests.

\end{document}